\documentclass[aps,prb,twocolumn,groupedaddress,showpacs,showkeys]{revtex4}
\usepackage{graphicx}

\input{tcilatex}
\begin{document}

\title{Quantum capacitor with discrete charge-anticharge: spectrum and
forces }
\author{J. C. Flores}
\affiliation{Instituto de Alta Investigaci\'{o}n IAI, Universidad de Tarapac\'{a},
Casilla 7-D Arica, Chile.}

\begin{abstract}
The quantum capacitor with discrete charge is modeled by \ a Hamiltonian
containing an inductive intrinsic term (tunnel effect between plates). The
spectrum is obtained using a double Hilbert space. Fluctuations in the
charge-anticharge pairs (zero total charge) give rise to an elementary
attraction which is compared to the Casimir force. In this case, the
field-fluctuations force could be also interpreted as charge-fluctuations
force.
\end{abstract}

\maketitle

\textbf{Introduction. -} \ Manifestations of charge discreteness in
nano-electronic \ (and mesoscopic) have\ been well established. For
instance, persistent electrical current phenomena in metallic coherent rings 
\cite{imry,buttiker} have a flux periodicity due to discrete charge
existence. Also, in experimental realizations \cite{santos,heiss,ventro} of
quantum dots, single charges are manipulated inside an entourage composed of
semiconductors associated to capacitances and gate voltages. In this sense,
quantum theory of electric circuits (with discrete charge) becomes an
important tool to understand a variety of electronic devices where the
length of quantum coherence is bigger than the size of the system.

Consider the quantum Hamiltonian of a LC circuit \cite{louisell} with
discrete charge \cite{li-chen,flores-epl,flores,utre} $q_{e}$:

\begin{equation}
\widehat{H}=\frac{2\hbar ^{2}}{L_{o}q_{e}^{2}}\sin ^{2}(\frac{q_{e}\widehat{%
\phi }}{2\hbar })+\frac{1}{2C_{o}}\widehat{q}^{2},  \label{H}
\end{equation}%
where $L_{o}$ is the \ inductance and $C_{o}$ the capacitance. The flux \ ($%
\widehat{\phi }$) and the charge operator ($\widehat{q}$) satisfies the
usual commutation rule $[\widehat{q},\widehat{\phi }]=i\hbar $. For
electrical circuits, it is more convenient to proceed via the time-evolution
equation for charge since physical measurements are related to electrical
current and charge directly in the respective elements (capacitances and
inductances). Moreover, the incorporation of an eventual electrical
resistance becomes operative through the evolution equation \cite{utre}.
Using Heisenberg equations \cite{bohm}, the temporal evolution for the
charge operator is given by

\begin{equation}
\frac{d^{2}}{dt^{2}}\widehat{q}=-\frac{1}{2C_{o}L_{o}}\left\{ \widehat{R}%
\widehat{q}+\widehat{q}\widehat{R}\right\} ,\text{ }  \label{evo}
\end{equation}

\[
\text{where\ }\widehat{R}=\sqrt{\left( 1-\left( \frac{L_{o}q_{e}}{\hbar }%
\frac{d\widehat{q}}{dt}\right) ^{2}\right) } 
\]

To clarify Eq. (\ref{evo}): \ (a) the deduction comes directly from the
dynamics equations $d\widehat{q}/dt=(\hbar /L_{o}q_{e})\sin (q_{e}\widehat{%
\phi }/\hbar )$ and \ $d\widehat{\phi }/dt=-\widehat{q}/C_{o}$. A first
integral of (\ref{evo}) gives directly the Hamiltonian (\ref{H}) with the
convenient choice for the integration constant. \ (b) The auxiliary variable 
$\widehat{R}$ corresponds directly to $\cos (q_{e}\widehat{\phi }/\hbar )$
expressed as function of the electrical current (the positive root is
compatible with oscillatory solutions).

In this work we deal with the solution of (\ref{evo}) which is a nonlinear
differential equation. Note that in the formal limit $q_{e}\rightarrow 0$
(strictly, $\phi \ll \hbar /q_{e}$) one obtains the well known linear
differential equation for the LC-circuit. In the next sections we found
explicitly a plane wave solution of (\ref{evo}) by working in the doubled
Hilbert space (charge-anticharge). The spectrum of the system is found
explicitly. The connection with the classical expression for the force
between the capacitor plates is obtained validating the idea that a quantum
capacitor has intrinsically associated an inductance (tunnel effect). A
comparison with the Casimir force is established.

\textbf{Charge-anticharge solution}. - We shall consider the equation (\ref%
{evo}) inside a doubled Hilbert space ($2\otimes \mathcal{H}$) and the time
dependent charge operator solution (frequency $\omega $):

\begin{equation}
\widehat{Q}_{t}=\widehat{q}\left( 
\begin{array}{cc}
0 & e^{i\omega t} \\ 
e^{-i\omega t} & 0%
\end{array}%
\right) ,  \label{solution}
\end{equation}%
where the hermitian operator $\widehat{q}$ (acting on the Hilbert space $%
\mathcal{H}$) is defined by the eigenvalues equation \cite{li-chen} in the
Schr\"{o}dinger picture: $\widehat{q}\left\vert n\right. \rangle
=nq_{e}\left\vert n\right. \rangle $, with $n$ integer, and related to
charge discreteness. Note that the solution (\ref{solution}) refers to the
charge in one plate, nevertheless, we assume total charge zero in the
capacitor.

From (\ref{solution}) \ one has the useful expressions

\begin{equation}
\left( \frac{d\widehat{Q}_{t}}{dt}\right) ^{2}=\widehat{q}^{2}\omega
^{2}\left( 
\begin{array}{cc}
1 & 0 \\ 
0 & 1%
\end{array}%
\right) ,\text{ and }\frac{d^{2}\widehat{Q}_{t}}{dt^{2}}=-\omega ^{2}%
\widehat{Q}_{t}  \label{curr}
\end{equation}

Putting the solution (\ref{solution}) \ into (\ref{evo}) (remember (\ref%
{curr})) and considering the channel $n$ of charge $nq_{e}$, one obtains the
implicit expression for the frequency $\omega _{n}$ for the circuit:

\begin{equation}
\omega _{n}^{2}=\frac{1}{L_{o}C_{o}}\sqrt{\left( 1-\left( \frac{%
L_{o}q_{e}^{2}}{\hbar }n\omega _{n}\right) ^{2}\right) },  \label{freq}
\end{equation}%
and, as expected, in the formal limit $q_{e}\rightarrow 0$ \ one has the
usual expression for the LC circuit with continuos charge. \ The figure 1
shows the structure of frequencies (\ref{freq}) of the system (normalized).
\ It is bounded and the maximum corresponds to the channel $n=0$. \ 

As in the case of the \textquotedblleft harmonic
oscillator\textquotedblright\ $d^{2}\widehat{q}/dt^{2}=-(1/L_{o}C_{o})%
\widehat{q}$, \ where the frequency $\omega _{o}^{2}=1/L_{o}C_{o}$ gives
origin to the a discrete spectrum ($E_{l}=\hbar \omega _{o}(l+\frac{1}{2})$%
), in this case we have the spectrum

\begin{equation}
E_{n,l}=\hbar \omega _{n}(l+\frac{1}{2})\text{ with }l=0,1,2,...
\label{spectrum}
\end{equation}%
Note that we have two quantum index ($n,l$) since we have doubled the space.
Here an interesting \ consequence must be mentioned, the diagonalization of
the two dimensional matrix in (\ref{solution}) is easy. Its eigenvalues are $%
\pm 1$ related to the couple charge-anticharge \ in the corresponding
solution of (\ref{evo}).

\textbf{Force due to charge fluctuations: comparison with Casimir. -}
Consider the background state $l=0$ in (\ref{spectrum}) and the channel $n=0$%
. That is, consider the energy state $E_{0,0}=\hbar /\left( 2\sqrt{L_{o}C_{o}%
}\right) $. As manifested before, the quantum capacitor has necessarily
associated \ an intrinsic inductance \ (tunnel effect). The relation between
electrical parameters is obtaned \ by imposing the equality between the
magnetic energy (for an elementary quantum flux) and the electrical energy
of the capacitor with elementary charge $q_{e}$. Namely, the intrinsic
inductance $L_{\hbar }$ associated to the quantum capacitor $C$ is given by

\begin{equation}
L_{\hbar }=\left( \frac{\hbar }{q_{e}^{2}}\right) ^{2}C.  \label{q-induc}
\end{equation}%
Putting this equation in the expression for the energy $E_{0,0}$, and
considering \ a parallel plate capacitor of area \ $A$ and separation $x$ it
becomes $E_{0,0}=\left( q_{e}^{2}x\right) /\left( 2\epsilon _{o}A.\right) $.
The force (by distance variation) between plates could then be written as \ 

\begin{equation}
F=-\frac{q_{e}^{2}}{2\epsilon _{o}A},  \label{for}
\end{equation}%
corresponding to the classical expression, but here the origin of the force
is related to charge-anticharge fluctuations in the plates. In this way,
equation (\ref{q-induc}) defines correctly the quantum inductance associate
to a quantum capacitor. It is a powerful result since (\ref{for}) validates (%
\ref{q-induc}). On the other hand, a circuit Hamiltonian could be expressed
as function of charges and currents or as function of the electromagnetic
fields. Classically for an inductance the magnetic field $B\sim \phi /S$ and
for a capacitor the electric field $E\sim Q/S$. This suggests that the
expression (\ref{for}) could be written as function of field properties.
Using the fundamental relation involving the structure fine constant $\alpha 
$ (i. e. $\alpha \hbar c4\pi \epsilon _{o}=q_{e}^{2}$, for electrons) \
equation \ (\ref{for}) could be written as

\begin{equation}
F=-(2\pi \alpha )\frac{1}{A}\hbar c.  \label{ftot}
\end{equation}

The so-called Casimir force \cite{casimir} $F_{cas}$ is related to frequency
contributions in the background energy of the electromagnetic fields with
explicit boundary conditions. For the parallel plate capacitor the Casimir
force is given by

\begin{equation}
F_{cas}=-\left( \frac{\pi ^{2}}{240}\right) \frac{A}{x^{4}}\hbar c,
\label{plaque}
\end{equation}%
and different from (\ref{ftot}). In fact, $F\ll F_{cas}$ because $x^{2}\ll A$
(corresponding to the usual condition for the capacitor). \ Note that the
Casimir force (related to fields) could be expressed as function of the
charge $q_{e}$ due to the\ structure fine constant existence. We will
exploit this point in the follow. Electrical circuits are by definition slow
time dependent systems \cite{landau}. Moreover, the Casimir effect
calculation considers the summation on low frequencies \ in the vacuum state
(regularization technique). In this way, a connection between both kind of
forces is expected. The connection becomes as

\begin{equation}
F_{cas}=N^{2}F,  \label{connect}
\end{equation}%
\ where $N\sim A/x^{2}$ must be interpreted directly as the number of
pair-fluctuations \ in the plates (and $Nq_{e}$ the total charge
fluctuations). So, the Casimir force could be interpreted as the force
between charge-anticharge fluctuations between the plates.

A final note, in the Casimir parallel plate problem, the physical quantities
\ (energy, force, etc.) are defined per unit of area. The only physical
length is the distance $x$. If we assume that the capacitor is composed of
elementary capacitors $C_{e}=\varepsilon _{o}x$ \ (in parallel) then $C=\sum
C_{e}$ or explicitly $\varepsilon _{o}A/x=N\varepsilon _{o}x$ and one
obtains $N=A/x^{2}$as conjectured before (i. e. every pair is associated
with an elementary capacitor).

\textbf{\ Conclusions and open questions. - \ }We have considered \ the
nonlinear charge evolution equation (\ref{evo}) corresponding \ to a LC
quantum circuit with discrete charge. \ The plane wave solution (\ref%
{solution}) pertains to a doubled Hilbert space (charge-anticharge). \ The
spectrum characterizing this kind of solution (\ref{spectrum}) has two
branches: one corresponding to \ the channel ($n$) and the other associated
with the usual harmonic oscillator structure ($l$). Since tunnel effects are
unavoidable there is an intrinsical inductance $L_{\hbar }$ (\ref{q-induc})
associated to the quantum capacitor. The force between plates (\ref{for}),
due to quantum charge fluctuations, was calculated and is smaller than the
corresponding Casimir force (\ref{plaque}). Nevertheless, from the
estimation of the number of fluctuating pairs $N=A/x^{2}$ this two
quantities become related (\ref{connect}). As pointed out in the
introduction, charge discreteness play an important role in mesoscopics.
Moreover, in NanoMENS devices \cite{santos} our description could be related
to transmission lines, quantum resonator, Coulomb blockade, quantum dots,
and others including quantum current magnification for two metallic rings 
\cite{floresprb}.

Finally \ there are three open questions. First, the frequency spectrum
decreases with the number of pair $n$\ (see figure 1) related possibly to
the fact that every pair is binding (negative energy). Second, for high
energy $E$ $\gg 2\hbar ^{2}/L_{o}q_{e}^{2},$ the electric part must be
dominant ($E\sim (nq_{e})^{2}/2C_{o}$). This corresponds to other kind of
solution different from (\ref{solution}). A third aspect, in reference \cite%
{flores-epl} it was mentioned that (\ref{H}) is not the only possible choice
to enforce the charge discreteness procedure. There are others flux periodic
Hamiltonian compatible with charge discreteness. In every case, one expect a
change in the spectrum in similar way as different crystal-cell composition
changes the band-gap structure. Nevertheless, expression (\ref{q-induc}) is
generic and does not depend on the particular form of the Hamiltonian.

\begin{center}
* * *
\end{center}

This work was supported by project FONDECYT\ \ 1070597. Valuable suggestions
by K. J. Chandia, M. Bologna and D. Laroze are acknowledges. Useful
discussions with A. P\'{e}rez concerning (\ref{q-induc}) are deeply
acknowledges.

Figure 1. The frequency (normalized) structure (\ref{freq}) of the system\
draws in continuos but the modes are indexed by integer values of the
variable $n$. \ When $n\rightarrow \pm \infty $ the frequency goes to zero.
That is, it diminishes with the number of binding pairs.

\FRAME{dtbpFUX}{3.1808in}{2.1205in}{0pt}{\Qcb{{}}}{}{Plot}{\special{language
"Scientific Word";type "MAPLEPLOT";width 3.1808in;height 2.1205in;depth
0pt;display "USEDEF";plot_snapshots TRUE;mustRecompute FALSE;lastEngine
"MuPAD";xmin "-5.102740344";xmax "5.102740344";ymin "-0.1012997669";ymax
"1.1066107669";xviewmin "-5.105800956";xviewmax "5.105800956";yviewmin
"-0.1016628392";yviewmax "1.1069708392";viewset"XY";rangeset"XY";plottype
12;labeloverrides 3;x-label "n";y-label "Freq";axesFont "Times New
Roman,12,0000000000,useDefault,normal";num-x-gridlines 24;num-y-gridlines
24;plotstyle "patch";axesstyle "normal";axestips FALSE;xis \TEXUX{x};yis
\TEXUX{y};var1name \TEXUX{$x$};var2name \TEXUX{$y$};function
\TEXUX{\EQN{6}{1}{}{}{\RD{\CELL{y^{2}=%
\sqrt{1-x^{2}y^{2}}}}{1}{}{}{}}};linecolor "black";linestyle 1;pointstyle
"point";linethickness 1;lineAttributes "Solid";var1range
"-5.102740344,5.102740344";var2range
"-0.1012997669,1.1066107669";num-x-gridlines 24;num-y-gridlines
24;curveColor "[flat::RGB:0000000000]";curveStyle "Line";VCamFile
'KRXQ6R02.xvz';valid_file "T";tempfilename
'KRXQ6R00.wmf';tempfile-properties "XPR";}}

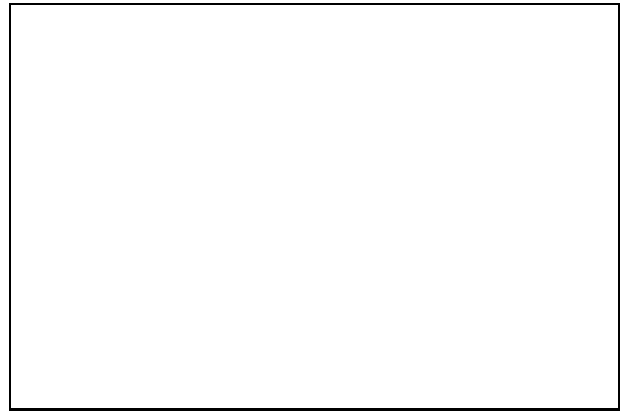
\end{document}